# What the Cited and Citing Environments Reveal of *Advances in Atmospheric Sciences*?

SHI Aolan[1,2] and Loet LEYDESDORFF[3]

[1]Institute of Atmospheric Sciences, Chinese Academy of Sciences, Beijing 100029
[2]School of Management, Graduate University of Chinese Academy of Sciences, Beijing 100190
[3] Amsterdam School of Communication Research (ASCoR), University of Amsterdam, Kloveniersburgwal 48, 1012 CX Amsterdam, The Netherlands; loet@leydesdorff.net; http://www.leydesdorff.net .

Abstract

The networking ability of journals reflects their academic influence among peer journals. This paper analyzes the cited and citing environments of the journal---*Advances in Atmospheric Sciences*---using methods from social network analysis. The journal has been actively participating in the international journal environment, but one has a tendency to cite papers published in international journals. *Advances in Atmospheric Sciences* is intensely interrelated with international peer journals in terms of similar citing pattern. However, there is still room for an increase in its academic visibility given the comparatively smaller reception in terms of cited references.

1. Introduction

At the era of information society, the exchanges of ideas become more intense and faster; there has been no time like the present one when there are so many different ways to communicate ideas. People can blog, twitter, email or the most traditional way---publish new ideas in some renowned journals. The latter approach is still the most common way for scientific researchers. The difference is that researchers nowadays are more concerned with the networking capability of the journals. This capacity is reflected by the indegree (cites to the journal) and outdegree (cites from the journal) citation activities.

Since Price (1965) first mentioned the concept of "delineating the topography of current scientific literature" based on "journal citations", researchers have been trying to find ways to optimize the maps. However, the journal-journal relationships are complex and the decomposition and aggregation of journal data has become a tough task (Leydesdorff, 1986, 2006). Leydesdorff (2007) listed the four most challenging problems as follows: 1) the choice of a seed journal, 2) the choice of similarity criteria, 3) the setting of threshold levels, and 4) the application of a clustering algorithm. Based on social network analysis, visualization of the citation impact environments of scientific journals were made available at the internet (e.g., at http://www.leydesdorff.net/jcr09) and the networking ability of journals can thus be delineated and animated (e.g., at http://www.leydesdorff.net/journals/nanotech).

The journal, *Advances in Atmospheric Sciences* (*Adv. Atmos. Sci.*) has devoted itself to the distribution of the newest achievements contributed by researchers in China and abroad. The journal has provided both a convenient and reliable platform for researchers to present their work and professional editing service (including proofreading, typesetting etc.). The latter have optimized the readability of the published papers (Wu et al., 2008). The journal's efforts have gained recognition. In May 2008, ScienceWatch.com named *Adv. Atmos. Sci.* a "Rising Star" among the Geosciences journals. According to the Essential Science Indicators of Thomson Reuters, the journal's current record includes 764 papers cited to a total of 1,658 times between 1 January 1998 and 29 February 2008 (Wu et al., 2008). Ten years after first having been indexed in *SCI-Expanded*, the journal is also included in core set of the *SCI* database since 2009.

While the journal has become one of the leading atmospheric journals in China, there is reason for concern about its academic influence worldwide. The journal's JCR impact factor fluctuates between 0.679 to 0.902 during the past three years (2007--2009 JCR reports). To find out the networking ability of the journal in the dynamic cited/citing environment and the role the journal has played among its peer international journals, the journal's citation maps were delineated for a more comprehensive analysis.

2. Method

Based on social network analysis, Leydesdorff's (2007) methods addressed the four problems as follows. First, the choice of the seed journal was brought under the hands of the end-users. As the purpose of research varies, the researcher can choose any journal as their focus and start the clustering of journals from there. The journals citing or cited by the focused journal will form the cited/citing environment of the journal.

The citing-cited relationship between journals forms a citation matrix. The cosine between the two vectors (Salton & McGill, 1983; Ahlgren et al., 2003) is chosen as the similarity measure between the distributions for the various journals included in a citation environment. Cosine values below 0.2 are suppressed to make the visualizations more focused and clear.

And last, the citation data of each journal is normalized by the number of citation divided by the numbers of total citation in the cited/citing environment. For example, a journal can be denoted as an ellipse in the cited environment (citing environment) of the seeding journal composed of all citing journals of (journals cited by) the seed journal. The length of the y-axis of the ellipse is the total indegree of cites to the journal (outdegree cites from the journal) divided by the total indegree (outdegree) citation numbers in the cited (citing) environment. The length of the x-axis of the journal is similar to that of y-axis, with the deduction of self-cites. For a more representative citation map, the threshold is set to 1% in the matrix to exclude the data of the journals whose cited/citing total are less than 1% of the cited/citing total of the seed journal.

On the one hand, the performance of a journal in the cited environment reflects its academic standing among peer journals: the impact breadth is reflected by the total number of citing journals and the impact depth is reflected by its quota in the citation map. On the other hand, the journal's performance in the citing environment reflects its tendency and vibrancy in the citation activities.

3. Analysis of *Adv. Atmos. Sci.*
3.1 The cited environment of *Adv. Atmos. Sci.*

Based on the 2009 Journal Citation Reports, with *Adv. Atmos. Sci.* as the seed journal, the cited environment is presented in Fig. 1.

From Fig. 1, it is easy to recognize that *J. Geophys. Res., Geophys. Res. Lett., J. Climate, Atmos. Environ.,* and *Month. Weather Rev.*, are the five most influential journals in the citation map; these journals occupy the core area of the map. As one of the rare validation studies, Bensman (2010) compared the survey results of the chemistry department staff of Louisiana State University and journal use frequency at the University of Illinois Chemistry Library, with impact factors and total cites, respectively. The author found that the correlations between "total cites" and the appreciations by users were significantly higher than the correlations between the impact factor and users' appreciation. Since journal impact factor fluctuates yearly and is prone to errors (miscalculation of citable items; unauthentic citations, i.e. citations to journal's Chinese version are counted to the journal's English version though literally they are different journals but with the same English title), one can argue that the total cites might be the more appropriate indicator of a journal's influence among its core journal aggregation. The ranking of the five most cited journals with C/N values is highly related with that of total cites ($r$=99.0%, Table 1). In this case, both the total cites and the C/N values represent the journals' influence.

Table 1. The five most cited journals in *Adv. Atmos. Sci.*'s cited environment.

|  | *J. Geophys. Res.* | *Geophys. Res. Lett.* | *J. Climate* | *Atmos. Environ.* | *Month. Weather Rev.* |
|---|---|---|---|---|---|
| Impact factor | 3.082 | 3.204 | 3.363 | 3.139 | 2.238 |
| Total cites | 144430 | 52131 | 20458 | 28524 | 15391 |
| C/N value | 40.866908 | 16.450265 | 11.37804 | 8.08881 | 7.431376 |

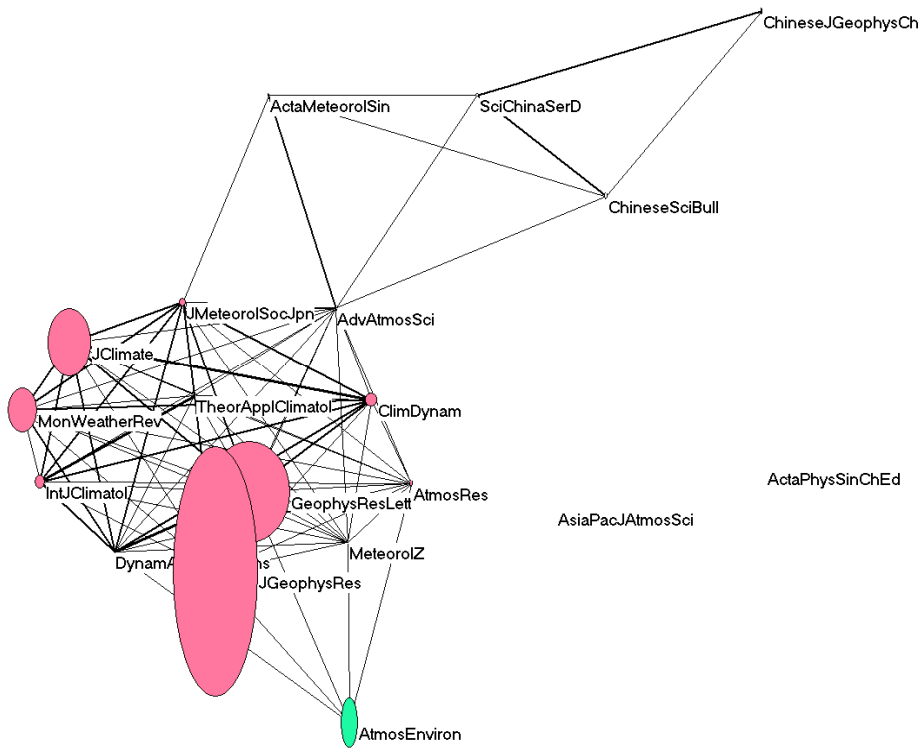

Fig. 1 The cited environment of *Adv. Atmos. Sci.*

Apart from the core, *Adv. Atmos. Sci.* with four other Chinese journals (*Acta Meteorl. Sin., Sci. China SerD, Chinese J. Geophys-CH, and Chinese Sci. Bull.*) are at the edge of the map, which shows that Chinese journals have comparatively low impact on peer journals internationally. The two journals, *Adv. Atmos. Sci.* and *Acta Meteorl Sin.*, function as bridges which connect the isolated Chinese journals with the international ones. Though *Chinese Sci. Bull* is a journal in general sciences, its realm of influence is on journals in geosciences (Zhou et al., 2005). Therefore, the five Chinese journals could be roughly classified as journals in geosciences. The map shows that though the Chinese journals have a comparatively weak international influence, they have formed a combining journal aggregation as a growing Chinese power. This is rare among other Chinese SCI journals.

Though *Acta Phys. Sinica* occupies a comparatively higher C/N rate (3.738837) in the map, its self cite rates is as high as 99.09% and its apparent no-citation relation with other journals exclude it from the journal aggregation groups. Table 2 teaches us that though *Acta Meteorl. Sin.* has a higher impact factor, *Adv. Atmos. Sci.* has a higher number of total cites and a higher C/N value, which means that *Adv. Atmos. Sci.* has another type of academic influence in this cited environment.

Table 2. The five Chinese journals in *Adv. Atmos. Sci.*'s cited environment.

|  | *Chinese J. Geophys-CH* | *Chinese Sci. Bull.* | *Sci China SerD* | *Adv. Atmos. Sci.* | *Acta Meteorl. Sin.* |
|---|---|---|---|---|---|
| Impact factor | 0.844 | 0.917 | 0.880 | 0.691 | 0.874 |
| Total cites | 1578 | 5116 | 2032 | 899 | 678 |
| C/N value | 1.159513 | 1.019589 | 0.602905 | 0.575127 | 0.488703 |

3.1 The citing environment of *Adv. Atmos. Sci.*

Based on the 2009 Journal Citation Reports, with *Adv. Atmos. Sci.* as the seed journal, the citing environment is presented in Fig. 2.

Similar to that of the cited environment, *J. Geophys. Res., Geophys. Res. Lett., J. Climate, Month. Weather Rev., J. Atmos. Sci.,* and *Climate Dyn.* are the six most active journals in this map. In contrast to the cited environment, however, *Adv. Atmos. Sci.* is comparatively active in the citing environment, which means that the authors in the journal are very sensitive to the research published in international journals and consider these journals as more reliable and direct sources of information.

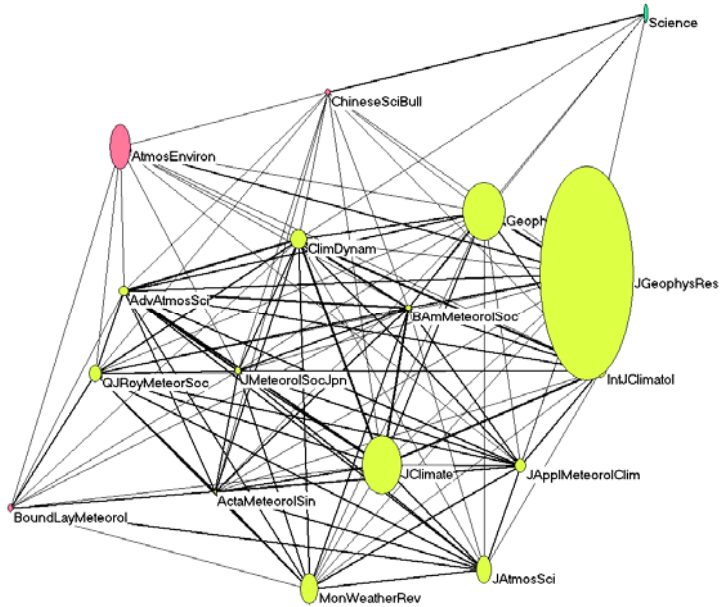

Fig. 2   The citing environment of *Adv. Atmos. Sci.*

For a better understanding of the citing behavior of each journal, we have elaborated the citing information of each journal in Table 3. The left column list the journal with their titles in the citing direction and the numbers 1-17 in row 1 represent the journals in the cited direction, respectively.

Table 3 The citing matrix of *Adv. Atmos. Sci.* as the seed journal

|  | 1 | 2 | 3 | 4 | 5 | 6 | 7 | 8 | 9 | 10 | 11 | 12 | 13 | 14 | 15 | 16 | 17 |
|---|---|---|---|---|---|---|---|---|---|---|---|---|---|---|---|---|---|
| *Acta Meteorl. Sin.* | 191 | 63 | 6 | 35 | 7 | 18 | 19 | 32 | 3 | 24 | 117 | 82 | 71 | 51 | 142 | 37 | 11 |
| *Adv. Atmos. Sci.* | 78 | 156 | 62 | 94 | 61 | 44 | 78 | 180 | 56 | 73 | 235 | 343 | 308 | 86 | 192 | 104 | 56 |
| *Atmos. Environ.* | 0 | 15 | 5216 | 70 | 119 | 18 | 16 | 722 | 20 | 193 | 114 | 54 | 2647 | 15 | 92 | 71 | 389 |

| Journal | | | | | | | | | | | | | | | | |
|---|---|---|---|---|---|---|---|---|---|---|---|---|---|---|---|---|
| B. Am. Meteorol. Soc. | 2 | 0 | 43 | 238 | 23 | 0 | 47 | 183 | 21 | 53 | 87 | 279 | 269 | 14 | 178 | 65 | 121 |
| Bound.-Lay. Meteorol. | 0 | 3 | 141 | 36 | 657 | 0 | 2 | 17 | 17 | 164 | 315 | 28 | 99 | 12 | 104 | 173 | 5 |
| Chinese Sci. Bull. | 28 | 31 | 59 | 19 | 0 | 544 | 25 | 97 | 0 | 0 | 42 | 37 | 255 | 23 | 28 | 0 | 483 |
| Climate Dynam. | 3 | 23 | 4 | 197 | 11 | 3 | 514 | 485 | 147 | 28 | 342 | 1127 | 474 | 71 | 366 | 209 | 228 |
| Geophys. Res. Lett. | 0 | 13 | 187 | 348 | 35 | 11 | 270 | 3150 | 136 | 104 | 621 | 1035 | 4651 | 101 | 327 | 200 | 1299 |
| Int. J. Climatol. | 5 | 14 | 33 | 211 | 22 | 3 | 163 | 374 | 685 | 86 | 125 | 811 | 350 | 37 | 244 | 92 | 124 |
| J. Appl. Meteorol. Clim. | 0 | 5 | 99 | 266 | 196 | 0 | 17 | 122 | 95 | 684 | 344 | 279 | 308 | 47 | 403 | 112 | 33 |
| J. Atmos. Sci. | 0 | 5 | 23 | 211 | 127 | 0 | 57 | 311 | 6 | 135 | 2832 | 443 | 612 | 120 | 708 | 542 | 99 |
| J. Climate | 0 | 51 | 15 | 768 | 16 | 4 | 640 | 1511 | 337 | 147 | 1358 | 4222 | 1415 | 284 | 1011 | 435 | 610 |
| J. Geophys. Res. | 0 | 62 | 1384 | 868 | 235 | 0 | 320 | 8117 | 279 | 643 | 1928 | 1753 | 26382 | 276 | 1098 | 725 | 2549 |
| J. Meteorol. Sco. Jpn. | 5 | 11 | 15 | 98 | 49 | 0 | 42 | 108 | 11 | 121 | 255 | 227 | 152 | 292 | 214 | 80 | 31 |
| Mon. Wea. Rev. | 4 | 11 | 6 | 406 | 53 | 0 | 54 | 196 | 35 | 221 | 1375 | 460 | 343 | 118 | 2581 | 696 | 50 |
| Q. J. Roy. Meteor. Soc. | 0 | 2 | 23 | 177 | 132 | 0 | 52 | 163 | 26 | 132 | 735 | 266 | 407 | 48 | 584 | 772 | 55 |
| Science | 0 | 0 | 20 | 0 | 0 | 0 | 25 | 227 | 0 | 0 | 0 | 75 | 437 | 0 | 0 | 0 | 3387 |

From this journal citation matrix, it can be inferred that the Chinese journals cite intensively the international ones but they do not receive citations in return to the same extent. The international journals are the sources and the Chinese journals relatively the sinks of citations.

There are different journal layers of citation concentration as revealed in Fig. 2. To find out the citing patterns of the journals, a "Principal Component Analysis" was applied as shown in Table 4.

A three-factor solution of the data matrix (explaining 64.9% of the variance) reveals that there are three groups with different citing patterns.

Table 4. Rotated Component Matrix of the citing journal matrix with *Adv. Atmos. Sci.* as the seeding journal (Extraction Method: Principal Component Analysis. Rotation Method: Varimax with Kaiser Normalization, and Rotation converged in 4 iterations）.

|  | Component | | |
|---|---|---|---|
|  | 1 | 2 | 3 |
| *Climate Dynam.* | 0.924 | 0.184 |  |
| *J. Climate* | 0.921 | 0.244 |  |
| *Int. J. Climatol.* | 0.86 |  |  |
| *Bound.-Lay. Meteorol.* | 0.759 | 0.307 | 0.377 |
| ***Adv. Atmos. Sci.*** | **0.742** | **0.521** | **0.218** |
| *Q. J. Roy. Meteor. Soc* | 0.141 | 0.856 |  |
| *J. Atmos. Sci.* |  | 0.819 |  |
| *Mon. Wea. Rev.* | 0.156 | 0.786 | -0.166 |
| *J. Meteorol. Sco. Jpn.* | 0.421 | 0.671 | -0.102 |
| *J. Appl. Meteorol. Clim.* | 0.138 | 0.649 |  |
| *Bound.-Lay. Meteorol.* | -0.421 | 0.513 |  |
| *J. Geophys. Res.* | 0.312 | 0.208 | 0.824 |
| *Geophys. Res. Lett.* | 0.466 | 0.138 | 0.799 |
| *Atmos. Environ.* | -0.141 |  | 0.601 |
| *Chinese Sci. Bull.* |  | -0.348 | 0.59 |
| *Acta Meteorl. Sin.* | 0.183 | 0.483 | -0.234 |
| *Science* |  | -0.292 | 0.481 |

*Climate Dynam., J. Climate, Int. J. Climatol., Bound.-Lay. Meteorol.,* and *Adv. Atmos. Sci.* have similar citing patterns. *Q. J. Roy. Meteor. Soc, J. Atmos. Sci., Mon. Wea. Rev.,*

*J. Meteorol. Sco. Jpn., J. Appl. Meteorol.,* and *Clim. Bound.-Lay. Meteorol.* form a second grouping with similar citing behavior. *J. Geophys. Res., Geophys. Res. Lett., Atmos. Environ.* and *Science* are more focusing on general geosciences in this citing environment and they have similar citing patterns.

The authors in *Adv. Atmos. Sci.* have actively cited publications in world first-class journals in atmospheric sciences and its similar citing behavior reveals the journal's intense interrelationship with the international ones. Though the journal has not received citations in proportion to what it has cited (Table 3), it has a wider academic influence internationally compared to other Chinese journals in this citing environment as only two of these eleven journals have not cited it.

4. Conclusion

Though *Adv. Atmos. Sci.* has become one of the leading journals in China, there is still room for increase in its international academic renown. Authors in the journal have an active tendency to cite papers published in international journals, and consequently, the journal's citing pattern is similar to that of comparable international journals.

The number of Chinese SCI journals has increased a lot. Totally, there are 114 Chinese journals indexed in SCI in 2009 JCR Report as compared to 75 in 2004. However, the international influence of Chinese journals is comparatively week as shown by the small numbers of citations received in the citation environment. In some fields, the influence of the Chinese SCI journals is growing in the form of journal aggregations. For example, the Chinese SCI journals in geosciences have formed a combining journal aggregation as a growing Chinese power. However, such action is exceptional among other Chinese SCI journals.

References

Bensman, S. J., 2010: Citations as Measures of Journal Importance: Total Citations Versus Impact Factor in Chemistry. *Journal of the American Society for Information Science* *and Technology*. (in press)

Leydesdorff, L., 1986: The development of frames of references. *Scientometrics,* **9,** 103-125.

Leydesdorff, L., 2006: Can scientific journals be classified in terms of aggregated journal-journal citation relations using the Journal Citation Reports? *Journal of the American Society for Information Science and Technology*, **57**(5), 601-613.

Leydesdorff, L., 2007: Visualization of the citation impact environments of scientific journals: An online mapping exercise. *Journal of the American Society for Information Science and Technology*, **58**, 25–38, doi: 10.1002/asi.20406.

Price, Derek J. de Solla, 1965: "Networks of Scientific Papers". *Science*, **149** (3683), 510–515.

Salton, G., and M. J. McGill, 1983: *Introduction to Modern Information Retrieval*. New York, McGraw Hill Book Co., 448pp.

Wu, G. X., H. J. Wang, and D. R. LÜ, 2008: Advances in Atmospheric Sciences---
A featured journal from Essential Science Indicators[SM]. [Available online from http://sciencewatch.com/ inter/jou/2008/08julAdvAtmospSci/]

Zhou, P., Loet Leydesdorff, and Y. S. Wu, 2005: Visualization of the citation environments of Chinese scientific and technological journals. *Chinese Journal of Scientific and Technical Periodicals*, 2005, 16(6), 773-780.